\documentclass[nofootinbib,twocolumn,aps,amsmath,amssymb,tightenlines]{revtex4}
\usepackage{graphics,bm}
\usepackage{epsfig}
\usepackage{graphicx}

\newcommand{\beq}{\begin{equation}}
\newcommand{\eeq}{\end{equation}}
\newcommand{\bqa}{\begin{eqnarray}}
\newcommand{\eqa}{\end{eqnarray}}

\begin{document}

\parindent=20pt
\parskip=10pt
\pagestyle{plain}
\font\tenrm=cmr10
\def\sumint{\hbox{$\sum$}\!\!\!\!\!\!\int}
\def\square{\vcenter{\vbox{\hrule height.4pt
          \hbox{\vrule width.4pt height4pt
          \kern4pt\vrule width.3pt}\hrule height.4pt}}}
\def\boxx{\square}
\def\isumint{\hbox{${\scriptstyle \Sigma}$}\!\!\!\!\!\int}
\def\isumdiff{\hbox{${\scriptstyle \Delta}$}\!\!\!\!\!\int}
\def\ranglec{\rangle_{\!\!c}}
\def\ranglex{\rangle_{\!\!x}}
\def\ranglecx{\rangle_{\!\!c,x}}

\title{Phase fluctuations in 
low-dimensional Gross-Neveu models}
\author{Jens O. Andersen} \affiliation{Nordita,\\
Blegdamsvej 17, DK-2100 Copenhagen {\O}, Denmark}

\date{\today}

\begin{abstract}
We consider the Gross-Neveu model with a continuous chiral symmetry
in two and three spacetime
dimensions at zero and finite temperature. In order to study
long-range phase coherence, we construct an effective 
low-energy Lagrangian for the
phase $\theta$. This effective Lagrangian is used to show that the
fermionic two-particle correlation function at finite temperature
decays algebraically in 2+1 dimensions and exponentially in 1+1 dimensions.

\end{abstract}

\email{jensoa@nordita.dk}

\maketitle

{\it Introduction.}
Low-dimensional field theories have been studied extensively as toy models
for QCD at zero and finite temperature. 
Many of these theories are remarkably rich, sharing a number of properties
with four-dimensional QCD. For example, both the Gross-Neveu model and 
the $O(N)$ nonlinear sigma model in 1+1 dimensions are 
asymptotically free and have nonperturbatively generated mass gaps.
In addition, the latter has instanton solutions for $N=3$.
In contrast with QCD in 3+1 dimensions, these models are renormalizable
in the $1/N$ expansion both in 1+1 and 2+1 dimensions.
Thus one can systematically study these models in 
a nonperturbative setting~\cite{nov,rosin}.
There is, however, an important difference between QCD and 
low-dimensional
field theories, namely the possibility of 
spontaneous breaking of continuous global symmetries. In QCD
with $N_f$ massless fermions, chiral symmetry is spontaneously broken at
zero temperature and this gives rise to $N_f^2-1$ massless Goldstone particles.
For $N_f=2$, these are the well known pions.
Chiral symmetry is expected to be restored at a critical temperature
of approximately 150 MeV depending on the number of flavors.
In 1+1 and 2+1 dimensions, the Mermin-Wagner theorem~\cite{mermin,coleman}
forbids the spontaneous breakdown of continuous symmetries at any finite 
temperature~\footnote{In 2+1 dimensions, spontaneous breaking of 
continuous symmetries may occur at $T=0$. 
Consequently, the critial temperature for
the restoration of a continuous symmetry is $T_c=0$.}. The reason is that the
phase fluctuations are so strong that they destroy the presence of a
condensate. The role of phase fluctuations was studied 
in the seminal papers on the Kosterlitz-Thouless phase transition of vortex 
unbinding~\cite{bere,thouless}.
The low-temperature phase where the vortices are bound in pairs 
is characterized by
an algebraic decay of the one-particle density matrix and 
superfluidity. The high-temperature phase of unbound vortices is
characterized by loss of superfluidity and exponential fall-off of 
correlations.
More recently, phase fluctuations have been studied in connection with 
Bose-Einstein condensation of atomic gases in harmonic traps
in one -and two-dimensional 
Bose gases~\cite{gor,utrecht1,utrecht2,utrecht3}.
In the following, we consider the Gross-Neveu model with a continuous
chiral symmetry~\cite{gv}. Many of its properties have been 
examined in detail~\cite{gv,rosin,edden,babaev,tc21,tc11}. In the present paper
we discuss the possible phases of the theory.
By using density and phase variables, one can conveniently
address the issues of long-range order and phase structure as a function of
temperature. 
These issues were discussed in Refs.~\cite{edden,babaev}; here we present
a somewhat more general treatment at zero and finite temperature.

{\it Gross-Neveu model.}
The Euclidean Lagrangian of the Gross-Neveu model with a $U(1)$
symmetry is~\cite{gv}
\bqa
\label{chiral}
{\cal L}&=&
\bar{\psi}_j\partial\!\!\!/\psi_j
-{g^2\over2N}\bigg[(\bar{\psi}_j\psi_j)^2-(\bar{\psi}_j\gamma_5\psi_j)^2
\bigg]\;,
\label{l0}
\eqa
where $j=1,2,...N$. In 2+1 dimensions, the
$\gamma$-matrices are $4\times4$ matrices and in 1+1 dimensions they
are $2\times2$ matrices~\cite{rosin}. 
They satisfy $\{\gamma_i,\gamma_j\}=2\delta_{ij}$ and
$\gamma_5=-i\gamma_0\gamma_1$.
The Lagrangian~(\ref{chiral}) has a $U(1)$ chiral symmetry:
\bqa
\psi_j\rightarrow e^{i\phi\gamma_5}\psi_j\;,
\eqa
where $\phi$ is a constant phase.
Following~\cite{gv}, we 
introduce the auxiliary fields $\sigma={g^2\over N}\bar{\psi}_j\psi_j$
and $\pi={g^2\over N}\bar{\psi}_ji\gamma_5\psi_j$.
The Lagrangian can then be written as
\bqa
{\cal L}&=&
\bar{\psi}_j\partial\!\!\!/\psi_j
+{N\over2g^2}\left(\sigma^2+\pi^2\right)
-\bar{\psi}_j\left(\sigma+i\pi\gamma_5\right)\psi_j
\;.
\label{l1}
\eqa
In terms of the fields $\sigma$ and $\pi$, chiral symmetry 
can be written as 
\bqa
\left(\sigma+i\pi\right)\rightarrow\left(\sigma+i\pi\right)e^{2i\phi}\;.
\eqa
Defining the density $\rho$ and phase $\theta$ variables by
\bqa
\sigma+i\pi&=&\rho e^{i\theta}\;,
\eqa
and the new fermion field $\chi$ by
\bqa
\chi_L=e^{-i\theta/2}\psi_L
\;,\hspace{1cm}
\chi_R=e^{i\theta/2}\psi_R\;,
\eqa
we can write the Lagrangian~(\ref{chiral}) as 
\bqa
{\cal L}&=&\bar{\chi}\left(
\partial\!\!\!/+{1\over2}i\gamma_5\partial\!\!\!/\theta+\rho
\right)\chi+{N\over2g^2}\rho^2\;.
\label{lag}
\eqa
In terms of $\rho$ and $\theta$, the chiral symmetry is simply 
$\theta\rightarrow\theta+c$, where $c$ is a constant. In these
variables the chiral symmetry is manifest.
Moreover, the operator $\partial\!\!\!/$ acting on $\theta$
in~(\ref{lag}) ensures that the field is massless and that it is
derivatively coupled.
The auxiliary field $\rho$ is now written as a sum of a space-time
independent background $\rho_0$~\footnote{Note that in 2+1 dimensions, 
$\rho_0$ is nonzero at $T=0$ 
only if the coupling $g$ is larger than a critical 
coupling 
$g_{\rm c}$~\cite{rosin}. In 1+1 dimensions, $\rho_0$ is always nonzero
at $T=0$.}
and a quantum fluctuating field 
$\tilde{\rho}$; $\rho=\rho_0+\tilde{\rho}$. 
Note that a nonzero expectation value for $\rho$
does not violate chiral symmetry since $\rho$
is invariant under global $U(1)$ transformations.
Since the Lagrangian~(\ref{lag}) is quadratic in the fermion fields, we
can integrate over $\chi$, and we obtain the effective action
for $\tilde{\rho}$ and $\theta$:
\bqa
S&=&\nonumber
-N{\rm Tr}\ln\left[
\partial\!\!\!/+{1\over2}i\gamma_5\partial\!\!\!/\theta+\rho_0+\tilde{\rho}
\right]
\\ &&
+{N\over2g^2}\int_0^{\beta}d\tau\int d^{d-1}x\;
(\rho_0+\tilde{\rho})^2\;,
\label{effact}
\eqa
where $\beta=1/T$ is the temperature (in units where $\hbar=1$) 
and $d-1$ is the dimension of space.
The trace is over both Dirac indices and spacetime.
The next step is to expand the functional determinant around the
classical solution of the equation of motion $\theta=0$ and $\tilde{\rho}=0$.
This expansion generates propagators for $\tilde{\rho}$
and $\theta$ as well as interaction vertices among them.

At leading order in $1/N$, the auxiliary fields $\rho$ and $\theta$ do
not propagate. However, at next-to-leading order, quantum fluctuations
induce the inverse propagators. These are obtained from expanding the
functional determinant in~(\ref{effact}) to second order in the fluctuations:
\bqa\nonumber
\Pi_{\rho}(Q)&=&{N\over g^2}-
{\rm Tr}(I)N\sumint_{\{P\}}
{1\over P^2+\rho_0^2}
\\ && \nonumber
+{1\over2}{\rm Tr}(I)N
\sumint_{\{P\}}
{4\rho_0^2+Q^2\over P^2+\rho_0^2}{1\over (P+Q)^2+\rho_0^2}\;,
\\ &&
\label{invrho}
\\\nonumber
\Pi_{\theta}(Q)&=&{1\over2}{\rm Tr}(I)N\rho_0^2
\sumint_{\{P\}}
{Q^2\over P^2+\rho_0^2}{1\over (P+Q)^2+\rho_0^2}\;,
\\ &&
\label{invtheta}
\eqa
where 
${\rm Tr}(I)$ is the trace of the identity matrix $I$.
We have introduced the notation
\bqa
\sumint_{\{P\}}&\equiv&T\!\!\!\!\!\!\!\!\!\!\!\!
\sum_{P_0=(2n+1)\pi T}\int{d^{d-1}p\over(2\pi)^{d-1}}\;,
\eqa
and $P=(P_0,{\bf p})$ is the Euclidean momentum. The integral over spatial
momenta is regularized by using an ultraviolet cutoff $\Lambda$.
Eq.~(\ref{invrho}) is ultraviolet divergent and the divergence can 
be eliminated by renormalizing the coupling constant $g$.
By analytic continuation to Minkowski space
($P_0\rightarrow i\omega+i\eta$) , one can
examine the analytic structure of the propagators for $\rho$ and $\theta$.
For example, at zero temperature, both propagators have a branch cut
starting at $\omega=\sqrt{q^2+4\rho^2_0}$. The $\theta$ propagator
has in addition a pole at $\omega=q$, which ensures that the field
is massless. This pole remains at finite temperature, which follows
immediately from Eq.~(\ref{invtheta}).

The effective potential ${\cal V}$ through leading order in $1/N$ is
\bqa
{\cal V}&=&{N\rho_0^2\over2g^2}-{1\over2}{\rm Tr}(I)
N\sumint_{\{P\}}
\ln\left[P^2+\rho_0^2\right]\;.
\label{effpot}
\eqa
The value of $\rho_0$ is determined by minizing the 
effective potential and to leading order in the $1/N$ expansion
this leads to the gap equation 
\bqa
{1\over g^2}&=&{\rm Tr}(I)\sumint_{\{P\}}{1\over P^2+\rho_0^2}\;.
\label{gap}
\eqa
This gap equation coincides with the gap equation in Gross-Neveu model
with a discrete $Z_2$ symmetry.
Eq.~(\ref{gap}) is ultraviolet divergent and it is 
rendered finite by renormalizing the coupling constant $g$.
At $T=0$, the gap equation can be solved explicitly for $\rho_0$. For example,
in 1+1 dimension, we obtain $\rho_0^2=\Lambda^2e^{-2\pi/g^2_R}$, 
where $g_R$ is the renormalized coupling and $\Lambda$ is the ultraviolet
cutoff. Finally, note that one can also 
use the gap equation~(\ref{gap}) to eliminate the divergent
sum-integral in the expression for the $\rho$-propagator.

If we are interested in the long-distance properties of the 
theory~(\ref{chiral}), we can simplify calculations by noticing
that for momenta $p$ much smaller than $\rho_0$, the massive field $\rho$
decouples and we are left with an effective long-distance theory for
the phase $\theta$. To leading order in derivatives, this can be written
as a free field theory:
\bqa
{\cal L}_{\rm eff}={1\over2}{\alpha}\left(\partial_{\mu}\theta\right)^2\;,
\label{long}
\eqa
where $\alpha$ is the stiffness of the phase fluctuations and is given by
\bqa
\alpha&=&{\rm Tr}(I)N\rho_0^2\sumint_{P}{1\over(P^2+\rho_0^2)^2}\;.
\label{stiff}
\eqa
The summation in~(\ref{stiff}) is over the bosonic Matsubara modes
$P_0=2\pi nT$. 
The expression for $\alpha$ follows immediately from~(\ref{invtheta}).
Chiral symmetry, $\theta\rightarrow\theta+c$, forbids terms in the effective
Lagrangian not involving derivatives of $\theta$ such as $\theta^4$
and $\theta^2\left(\nabla\theta\right)^2$. Thus the next operator
in the low-energy expansion involves four derivatives and the
results concerning the infrared properties of the Gross-Neveu model
obtained using~(\ref{long}) are stable against 
perturbations~\cite{edden}.

In order to discuss the possible phases of the Gross-Neveu model, we
must examine the long-distance behavior of the following four-fermion
correlation function~\cite{edden}
\bqa\nonumber
\rho_2(x,0)&=&
\left\langle
\bar{\psi}(x)(1+\gamma_5)\psi(x)
\bar{\psi}(0)(1-\gamma_5)\psi(0)
\right\rangle\;.
\\ &&
\eqa
The function $\rho_2(x,0)$ can be written as 
\bqa\nonumber
\rho_2(x,0)&=&\left\langle\rho(x)e^{-i\theta(x)}\rho(0)e^{i\theta(0)}
\right\rangle\\
&\approx&\rho_0^2e^{-{1\over2}\langle[\theta(x)-\theta(0)]^2\rangle}\;,
\eqa
where we in the second line
have replaced $\rho$ by its expectation value $\rho_0$
and used
Wick's theorem to rewrite the exponent.
This replacement is justified since density fluctuations are negligible 
for long distances.
Using the effective Lagrangian~(\ref{long}), the exponent~(\ref{exp})
can be written as
\bqa\nonumber
\left\langle[\theta(x)-\theta(0)]^2\right\rangle&=&{1\over\alpha}
\int{d^{d-1}p\over(2\pi)^{d-1}}\coth\left({\beta p\over2}\right)
\\ &&
\times
{\left[1-\cos({\bf p}\cdot {\bf x})\right]\over p}\;.
\label{exp}
\eqa
The behavior of~(\ref{exp}) in the limit $x\rightarrow\infty$ then determines
the possible phases of the Gross-Neveu model.

{\it 2+1 dimensions.}
We first notice that the gap equation~(\ref{gap}) and 
the stiffness~(\ref{stiff}) can be
calculated explicitly:
\bqa
{1\over g^2_R}&=&-{\rho_0\over\pi}-{2T\over\pi}
\ln\left[1+e^{-\beta\rho_0}\right]\;,\\ 
\alpha&=&{N\over2\pi}\rho_0\tanh{\beta\rho_0\over2}\;.
\eqa
Averaging over the angle between ${\bf p}$ and ${\bf x}$ in~(\ref{exp}), 
we obtain
\bqa\nonumber
\left\langle[\theta(x)-\theta(0)]^2\right\rangle&=&{1\over2\pi\alpha}
\int_{0}^{\infty}{dp}\coth\left({\beta p\over2}\right)
\Big[1-J_0(px)\Big]\;,
\label{exp2}\\ &&
\eqa
where $J_0(px)$ is a Bessel function of the first kind.
At $T=0$, the expression~(\ref{exp2}) reduces to
\bqa
\langle[\theta(x)-\theta(0)]^2\rangle&=&-{1\over2\pi\alpha x}\;.
\eqa
In the large-$x$ limit, the exponent vanishes
and this
shows that there is a real condensate for $T=0$. This is the usual
phase with the $U(1)$ symmetry spontaneously broken
and long-range order~\cite{rosin}.
At finite temperature 
the exponent~(\ref{exp2}) can not be evaluated in closed form.
However, one can expand the integrand and integrate term by term. This gives
the following convergent series
\bqa\nonumber
\langle[\theta(x)-\theta(0)]^2\rangle&=&-{1\over2\pi\alpha x}
\\ &&
\hspace{-1cm}
+{T\over\pi\alpha}\sum_{n=1}^{\infty}{1\over n}\left[
1-\sqrt{\beta^2n^2\over\beta^2n^2+x^2}
\right]
\;.
\label{n}
\eqa 
The large-$x$ behavior can be found by integrating the second term 
in~(\ref{n}) considering $n$ as a continuous variable.
Alternatively, one can the apply dimensional-reduction argument
that for distance scales $x$ much larger than $1/T$ and
$1/\rho_0$, the correlator is dominated by the zeroth Matsubara mode.
Either way, in the limit $x\rightarrow\infty$
one obtains
\bqa
\langle[\theta(x)-\theta(0)]^2\rangle&=&{T\over\alpha\pi}
\left[\ln{x\over\beta}+C\right]\;,
\eqa
where $C$ is a constant.
The correlator decreases algebraically with exponent $\gamma=T/2\pi\alpha$. 
This shows
that at finite temperature there is a quasicondensate whose density is
$\rho_0$. In this phase, there is a local gap, but the phase is incoherent.
This is referred to as a pseudogap.

The gap equation~(\ref{gap}) 
ceases to have a nontrivial solution for $\rho_0$
for temperatures $T$ above 
$T_{\rho}=\rho_0(T=0)/2\ln2$~\cite{rosin,tc21}
(Note that $T_{\rho}$
is the critical temperature for the phase transition
in the case of the 2+1 dimensional Gross-Neveu model with a $Z_2$-symmetry).
Above this temperature,
the leading-order effective potential then reduces
to the ideal-gas value, ${\cal V}=-3T^3\zeta(3)/4\pi$.

{\it 1+1 dimensions.}
Averaging over the angle between ${\bf p}$ and ${\bf x}$ in~(\ref{exp}), 
one obtains
\bqa\nonumber
\left\langle[\theta(x)-\theta(0)]^2\right\rangle&=&{1\over2\pi\alpha}
\int_0^{\infty}{dp}\coth\left({\beta p\over2}\right)
\\ &&\times
{\left[1-\cos(px)\right]\over p}\;,
\label{exp4}
\eqa
At zero temperature and in the large-$x$ limit, the exponent reduces to
\bqa
\langle[\theta(x)-\theta(0)]^2\rangle&=&{1\over2\pi\alpha}\ln x\;.
\eqa
At zero temperature, the stiffness~(\ref{stiff}) reduces to $\alpha=N/2\pi$
which implies that $\rho_2(x,0)=\rho_0^2x^{-1/N}$. This agrees with
Refs.~\cite{edden,rosin}.
Thus at $T=0$, the two-particle correlation function decays algebraically
and one has ``almost long-range order''.
At finite temperature 
the exponent~(\ref{exp4}) can be calculated exactly and reads
\bqa
\langle[\theta(x)-\theta(0)]^2\rangle&=&{1\over2\pi\alpha}\left[
\ln\sinh(\pi x/\beta)+K\right]\;,
\eqa
where $K$ is a constant.
We then obtain
\bqa
\lim_{x\rightarrow\infty}
\langle[\theta(x)-\theta(0)]^2\rangle&=&{1\over2\alpha}Tx\;,
\label{lx}
\eqa
which shows that the correlator decays exponentially,
$\rho_2(x,0)\sim e^{-{1\over4\alpha}Tx}$.
Thus not even a quasicondensate is present at finite temperature.
The large-$x$~(\ref{lx}) behavior could also have been derived by calculating 
the contribution from the static Matsubara mode, which dominates in this limit.

For temperatures above 
$T_{\rho}=\rho_0(T=0)e^{\gamma_E}/\pi$~\cite{tc11}, the
gap equation has only the trivial solution $\rho_0=0$
(The critical temperature for the phase transition in the discrete case
is $T_c=0$). 
The
leading-order effective potential therefore reduces to the ideal-gas
value, ${\cal V}=-\pi T^2/6.$ More generally, due to asymptotic freedom we
expect in the high-temperature limit that the $1/N$ expansion reduces to
perturbation theory. 

{\it Summary.}
In the present paper, we have studied the phase fluctuations of
the $U(1)$-symmetric Gross-Neveu model in 1+1 and 2+1 dimensions.
Constructing an effective low-momentum field theory for the phase $\theta$,
we have calculated the two-particle correlation function.
In 2+1 dimensions, there is a real gap at zero temperature, while the phase
fluctuations turn it into a pseudo-gap at finite temperature.
In 1+1 dimensions, the quasicondensate which is present at $T=0$, is
destroyed at finite temperature due to phase fluctuations.
In particular, this implies that the Kosterlitz-Thouless-like phase
present at $T=0$~\cite{edden} disappears at finite temperature.


\end{document}